\documentclass[preprint2]{aastex}

\begin{document}
\def \Angs        {\hbox{\rm \AA}}
\def \Vvec        {\hbox{$\mathbf V$}}
\def \nvec        {\hbox{$\mathbf {\hat e}_n$}}
\def \zvec        {\hbox{$\mathbf {\hat e}_z$}}
\def \OTSH        {\hbox{\rm OTSh}}
\def \VW          {\hbox{$v_{\rm w}$}}
\def \VWpara      {\hbox{$v_{{\rm w},\parallel}$}}
\def \VWperp      {\hbox{$v_{{\rm w},\perp}$}}
\def \VP          {\hbox{$v_P$}}
\def \V8          {\hbox{$v_8$}}
\def \Vrel          {\hbox{$v_{\rm rel}$}}
\def \TA          {\hbox{$T_{\rm A}$}}
\def \NNc          {\hbox{$\cal N$}}
\def \Tmax          {\hbox{$T_{\rm Max}$}}
\def \EMA          {\hbox{$EM_{\rm A}$}}
\def \HH          {\hbox{$\cal H$}}
\def \Rbase          {\hbox{$R_{\rm base}$}}
\def \Rclump          {\hbox{$R_{\rm clump}$}}
\def \jem          {\hbox{$j_{\rm em}$}}
\def \VPpara      {\hbox{$v_{P,\parallel}$}}
\def \VPperp      {\hbox{$v_{P,\perp}$}}
\def \Ain      {\hbox{$A_{1}$}}
\def \Aex      {\hbox{$A_{2}$}}
\def \DL       {\hbox{$\Delta \ell$}} 

\def \lam      {\hbox{$\lambda$}}
\def \Llam      {\hbox{$L_\lambda$}}
\def \wlam      {\hbox{$w_\lambda$}}
\def \kaplam    {\hbox{$\kappa_\lambda$}}
\def \Tw      {\hbox{$T_w$}}
\def \gamav   {\hbox{$\gamma_{\rm v}$}}
\def \gamaF   {\hbox{$\gamma_F$}}
\def \Rc   {\hbox{$R_c$}}
\def \Dclump   {\hbox{$D_c$}}
\def \dellam  {\hbox{$\delta \lambda$}}
\def \delw    {\hbox{$\delta w$}}
\def \flam    {\hbox{$f_\lambda$}}
\def \Flam    {\hbox{$F_\lambda$}}
\def \lamflam {\hbox{$\lambda f_\lambda$}}
\def \Lamlam {\hbox{$\Lambda_{\lambda}$}}
\def \gammaf  {\hbox{$\gamma_F$}}
\def \gammaw  {\hbox{$\gamma_W$}}
\def \taulam    {\hbox{$\tau_\lambda$}}
\def \tauff    {\hbox{$\tau_{\rm ff}$}}
\def \HeII{\hbox{He~{\sc ii}}}
\def \Stromgren       {\hbox{Str\"{o}mgren}}
\newcommand{\Halpha}{\mbox{H$\alpha$}}
\newcommand{\kmsec}{\mbox{$\rm{km \; s^{-1}}$}}
\newcommand{\Lnu}{\mbox{$L_\nu$}}
\newcommand{\EM}{\mbox{$EM$}}
\newcommand{\EMT}{\mbox{$EM(T)$}}
\newcommand{\Lstar}{\mbox{$L_*$}}
\newcommand{\jnu}{\mbox{$j_\nu$}}
\newcommand{\mH}{\mbox{$m_{\rm H}$}}
\newcommand{\mstar}{\mbox{$M_*$}}
\newcommand{\msun}{\mbox{$M_\odot$}}
\newcommand\Rsun{\hbox{$R_\odot$}}
\newcommand\Lsun{\hbox{$L_\odot$}}
\newcommand\Msun{\hbox{$M_\odot$}}
\newcommand\Rstar{\hbox{$R_\ast$}}
\newcommand\Tstar{\hbox{$T_*$}}
\newcommand\Teff{\hbox{$T_{eff}$}}
\newcommand\TL{\hbox{$T_L$}}
\newcommand\TX{\hbox{$T_X$}}
\newcommand\THHe{\hbox{$T_{HHe}$}}
\newcommand\TGR{\hbox{$T_{GR}$}}
\newcommand\hhe{\hbox{$H-He$}}
\newcommand\HWHM{\hbox{$HWHM$}}
\newcommand\EMX{\hbox{$EM_{\rm X}$}}
\newcommand\fir{\hbox{$fir$}}
\newcommand\eg{\hbox{e.g.,}}
\newcommand\ie{\hbox{i.e.,}}
\newcommand\etal{\hbox{et~al.}}
\newcommand\ROSAT{\hbox{\it ROSAT}}
\newcommand\ASCA{\hbox{\it ASCA}}
\newcommand\XMM{\hbox{\it XMM-Newton}}
\newcommand\Chandra{\hbox{\it Chandra}}
\newcommand\Copernicus{\hbox{\it Copernicus}}
\newcommand\Einstein{\hbox{\it Einstein}}
\newcommand\EUVE{\hbox{\it EUVE}}
\newcommand\Rlow{\hbox{$R_{\rm low}$}}
\newcommand\Rmin{\hbox{$R_{\rm min}$}}

\newcommand\CIV{\hbox{C {\sc iv}}}
\newcommand\NVII{\hbox{N {\sc vii}}}
\newcommand\NVI{\hbox{N {\sc vi}}}
\newcommand\NV{\hbox{N {\sc v}}}
\newcommand\OVI{\hbox{O {\sc vi}}}
\newcommand\OVII{\hbox{O {\sc vii}}}
\newcommand\OVIII{\hbox{O {\sc viii}}}
\newcommand\MgXI{\hbox{Mg {\sc xi}}}
\newcommand\MgXII{\hbox{Mg {\sc xii}}}
\newcommand\FeXVII{\hbox{Fe {\sc xvii}}}
\newcommand\NeIX{\hbox{Ne {\sc ix}}}
\newcommand\NeX{\hbox{Ne {\sc x}}}
\newcommand\SiIV{\hbox{Si {\sc iv}}}
\newcommand\SiXIII{\hbox{Si {\sc xiii}}}
\newcommand\SiXIV{\hbox{Si {\sc xiv}}}
\newcommand\SXV{\hbox{S {\sc xv}}}
\newcommand\SXVI{\hbox{S {\sc xvi}}}
\newcommand\ArXVII{\hbox{Ar {\sc xvii}}}
\newcommand\CaXIX{\hbox{Ca {\sc xix}}}
\newcommand\CaXX{\hbox{Ca {\sc xx}}}
\newcommand\CaXXI{\hbox{Ca {\sc xxi}}}
\newcommand\FeXX{\hbox{Fe {\sc xx}}}
\newcommand\FeXXII{\hbox{Fe {\sc xxii}}}
\newcommand\FeXXIII{\hbox{Fe {\sc xxiii}}}
\newcommand\FeXXIV{\hbox{Fe {\sc xxiv}}}
\newcommand\FeXXV{\hbox{Fe {\sc xxv}}}
\newcommand\ftoi{\hbox{$f/i$}}
\newcommand\PV{\hbox{P{\sc v}}}
\newcommand\HtoHe{\hbox{$H/He$}}
\newcommand\RRR{\hbox{\sf R}}
\newcommand\zpup{\hbox{$\zeta\ ${\rm Pup}}}
\newcommand\zori{\hbox{$\zeta\ ${\rm Ori}}}
\newcommand\zoriA{\hbox{$\zeta\ ${\rm Ori A}}}
\newcommand\COB{\hbox{Cyg OB2}}
\newcommand\cygate{\hbox{Cyg OB2 No. 8a}}
\newcommand\cygnin{\hbox{Cyg OB2 No. 9}}
\newcommand\dori{\hbox{$\delta\ ${\rm Ori}}}
\newcommand\doriA{\hbox{$\delta\ ${\rm Ori A}}}
\newcommand\tsco{\hbox{$\tau\ ${\rm Sco}}}
\newcommand\thoriC{\hbox{$\theta^1$ Ori C}}
\newcommand\eori{\hbox{$\epsilon\ ${\rm Ori}}}
\newcommand\ecma{\hbox{$\epsilon\ ${\rm CMa}}}
\newcommand\bcma{\hbox{$\beta\ ${\rm CMa}}}
\newcommand\sori{\hbox{$\sigma\ ${\rm Ori}}}
\newcommand\iori{\hbox{$\iota\ ${\rm Ori}}}
\newcommand\zoph{\hbox{$\zeta\ ${\rm Oph}}}
\newcommand\gcas{\hbox{$\gamma\ ${\rm Cas}}}
\newcommand\bcru{\hbox{$\beta\ ${\rm Cru}}}
\newcommand\xper{\hbox{$\xi\ ${\rm Per}}}
\newcommand\Berghofer{\hbox{Bergh\"{o}fer}}
\newcommand\vinf{\hbox{$V_\infty$}}
\newcommand\Mdot{{\hbox{$\dot M$}}}
\newcommand\Ndot{{\hbox{$\dot N$}}}
\newcommand\Msunyr{\hbox{$M_\odot\,$yr$^{-1}$}}
\newcommand\kms{\hbox{km$\,$s$^{-1}$}}
\newcommand\LXLBol{{\hbox{$L_X/L_{Bol}$}}}
\newcommand\Ne{\hbox{$N_e$}}
\newcommand\Np{\hbox{$N_p$}}
\newcommand\lamz{\hbox{$\lambda_{\circ}$}}
\newcommand\linflux{\hbox{line flux$/10^{-13}$}}
\newcommand\censhif{\hbox{$\frac{\displaystyle{\rm
centroid}}{\displaystyle{\rm shift}}$}}
\newcommand\shoceff{\hbox{$\frac{\displaystyle{\rm shock}}{\displaystyle{\rm
efficiency}}$}}
\newcommand{\n}[1]{\tablenotemark{\#1}}
\newcommand{\vsini}{$v \sin i$}
\newcommand\Rfir{\hbox{$R_{fir}$}}
\newcommand\Rtau{\hbox{$R_{\tau = 1}$}}
\newcommand\LB{{\it line-blending effects}}


\shorttitle{X-ray Spectra from Wind Clump Bow Shocks}

\title{X-ray Emission Line Profiles from Wind Clump Bow Shocks in
Massive Stars }

\author{R.~Ignace,}
\affil{
Department of Physics \& Astronomy, 
East Tennessee State University, Johnson City, TN, 37614, \\ ignace@etsu.edu}

\author{W.~L.~Waldron,}
\affil{
Eureka Scientific Inc., 2452 Delmer St., Oakland, CA, 94602,  \\
wwaldron@satx.rr.com}

\author {J.~P.~Cassinelli,}
\affil{
Department of Astronomy, University of Wisconsin-Madison, Madison, WI 53711,\\
cassinelli@astro.wisc.edu}

\author{A.~E.~Burke}
\affil{
990 Washington Street \#317,
Dedham, MA 02026 \\
burke.alexander@gmail.com
}

\begin{abstract}

The consequences of structured flows continue to be a pressing topic
in relating spectral data to physical processes occurring in massive
star winds.  In a preceding paper, our group reported on hydrodynamic
simulations of hypersonic flow past a rigid spherical clump to
explore the structure of bow shocks that can form around wind clumps.
Here we report on profiles of emission lines that arise from such
bow shock morphologies.  To compute emission line profiles, we adopt
a two component flow structure of wind and clumps using two ``beta''
velocity laws. While individual bow shocks tend to generate double
horned emission line profiles, a group of bow shocks can lead to
line profiles with a range of shapes with blueshifted peak emission
that depends on the degree of X-ray photoabsorption by the interclump
wind medium, the number of clump structures in the flow, and the
radial distribution of the clumps.  Using the two beta law prescription,
the theoretical emission measure and temperature distribution
throughout the wind can be derived.  The emission measure tends to
be power law, and the temperature distribution broad in terms of
wind velocity.  Although restricted to the case of adiabatic cooling,
our models highlight the influence of bow shock effects for hot
plasma temperature and emission measure distributions in stellar
winds and their impact on X-ray line profile shapes.  Previous
models have focused on geometrical considerations of the clumps and
their distribution in the wind.  Our results represent the first
time that the temperature distribution of wind clump structures are
explicitly and self-consistently accounted in modeling X-ray line
profile shapes for massive stars.

\end{abstract}


\section{Introduction} 

The subject of X-ray production in massive star winds continues to
be an evolving field of study.  The superionization seen at
UV~wavelengths of OB stars were best explained by a model that had
a source of X-rays in the winds (Cassinelli, Castor, \& Lamers 1978;
Cassinelli \& Olson 1979). Initial observations by the {\em Einstein}
observatory made the important discovery that essentially all O~stars
were X-ray sources (Harnden \etal\ 1979; Seward \etal\ 1979).  A
key finding to emerge from these early observations is that the
observed X-ray luminosities are roughly correlated with the bolometric
luminosities as $L_X\approx 10^{-7} L_{\rm Bol}$ (e.g., Cassinelli
\etal\ 1981).  Additional more extensive studies confirmed the
relationship (e.g., Berghofer \etal\ 1997; Naz\'{e} \etal\ 2011),
although the basis of the relationship continues to be a point of
investigation (e.g., Owocki \& Cohen, 1999; Owocki \etal\ 2011).
In addition, the majority of OB~stars display soft X-ray emissions
with temperatures $kT < 1$ keV (e.g., Berghofer \etal\ 1996; G\"{u}del
\& Naz\'{e} 2009);

Two pictures for the X-ray emission from hot stars arose:  one
involving a coronal zone at the base of a cool wind (Cassinelli \&
Olson 1979) and one involving shocks that form by line-driven wind
instabilities (Lucy \& White 1980; Lucy 1982).  The coronal model
as the sole source of the observed X-ray emission was quickly ruled
out based on analyses of the earliest higher spectral resolution
observations using the Solid State Spectrometer (SSS) on the {\em
Einstein} observatory.  Cassinelli \& Swank (1983) found that the
predicted large X-ray optical depths expected for a base coronal
source of X-rays were incompatible with the observed SSS spectra.
They further suggested that these winds consist of many shock
fragments to explain the lack of significant X-ray variability.

Studies of X-ray emissions from OB~stars have focused primarily on
exploring the wind driven instabilities (or line de-shadowing
instability, LDI) as a process of producing a distribution of wind
shocks (e.g., Owocki, Castor, \& Rybicki 1988). A detailed
picture of the expected X-ray production from these wind shocks was
given by Feldmeier (1995), and Felmeier \etal\ (1997) showed that
a wide range of temperatures could be produced in a planar shock front.

We are now in an era of high spectral resolution X-ray astronomy
with a few dozen massive stars having been studied in long pointed
observations (e.g., Walborn, Nichols, \& Waldron 2009).  
Better quality data have led to a host of new questions
concerning the physics of X-ray generation in massive star winds
(e.g., Waldron \& Cassinelli 2007; hereafter WC07).  Most of the
X-ray line emission is clearly formed within the winds. A triad of
lines from He-like ions (forbidden, intercombination, and resonance
or ``fir'' lines) provide direct information about the formation
radius of X-ray line emission (Kahn \etal\ 2001; Waldron \& Cassinelli
2001; Leutenegger \etal\ 2006). Supergiant winds typically show
that the lower energy ion stages such as \OVII\ tend to form near
or above 10~\Rstar; intermediate energy ions (e.g., \NeIX\ and
\MgXI) form deeper at $\approx 3$ to 8~\Rstar; and high energy ions
such as \SiXIII\ and \SXV\ form relatively close to the star ($<
2 \Rstar$).  Waldron \& Cassinelli (2001) suggested that these
differences in depths could perhaps be explained from considerations
of wind absorption effects, since the cool wind opacity scales as
$\kappa \propto \lambda^3$.  Thus winds are more transparent at
shorter wavelengths (higher energies). Waldron \& Cassinelli (2001)
also noticed that the location of line formation for the He-like
ions appeared to correlate with the respective radii of optical
depth unity for the X-ray photoabsorption (c.f., Cassinelli \etal\
2001; Miller \etal\ 2002; Oskinova \etal\ 2006; WC07).  The conclusion
is that hot plasma is spatially distributed in the wind
flow.

One surprising result from the high resolution X-ray spectroscopy
data is the general symmetry of broad lines and the frequent absence
of line profiles with significantly blue-shifted peak emissions.  It had been
expected that the wind X-ray lines would be generally broad yet
skewed to the short wavelength (or ``blueward'') side of the line.
This skewness is a consequence the fact that for an expanding wind,
the column depth of photoabsorption to the flow on the far side of
the star is larger than on the near side, resulting in differential
attenuation between the red and blueshifted hemispheres (MacFarlane
\etal\ 1991; Ignace 2001; Owocki \& Cohen 2001).  So the observation
of frequently symmetric and unshifted lines was unexpected for the
massive winds of OB~supergiants. For these stars WC07 found that
$\approx 60\%$ of emission lines are broad with a mean line-width
(HWHM) of 0.3 to 0.5 of the wind terminal speed (\vinf), and in
excess of 75\% of the lines have line-shifts that lie within $\pm0.2
\vinf$ of line center.

As suggested by Waldron \& Cassinelli (2001), the simplest way to
account for the rather symmetric and unshifted X-ray line profiles
is that the wind is more optically thin to X-rays than suggested
by the mass-loss rates.  A variety of models have emerged to explain
the line symmetry problem by studying wind clumping and wind porosity
effects.  Clumping in dense Wolf-Rayet winds has been known for
many years (Moffat \etal\ 1988), and there is direct evidence of
clumping among some O-stars (e.g., Lepine \& Moffat 2008). Clumping
can be categorized as ranging from micro-clumping (e.g., Hillier
1991; Hamann \& Koesterke 1998) to macro-clumping (e.g., Feldmeier
\etal\ 2003; Brown \etal\ 2004; Oskinova \etal\ 2004, 2006, 2007;
Owocki \& Cohen 2006), or a mix of the two.  Micro-clumping explicitly
assumes all clumps are optically thin at all wavelengths, which
need not be the case for macro-clumping.

Reductions in the mass-loss rate \Mdot\ by a factor of 10 or more
appeared to be supported by {\em FUSE} observations of \PV\ lines
from several hot-stars (Fullerton \etal\ 2006). Although this would
certainly make the winds more thin to X-rays, this severe reduction
in \Mdot\ can be eliminated either by accounting for wind
``macro-clumping'' (Oskinova \etal\ 2007) or by including the effects
of XUV radiation in reducing the fractional abundance of \PV\
(Waldron \& Cassinelli 2010).

Other models that have been proposed to explain the symmetry of the
X-ray lines include the effects of resonance line scattering on
line shapes (Ignace \& Gayley 2002), and there is support in one
case where such effects are applicable (Leutenegger \etal\ 2007);
two-component wind structures where the polar wind component is
impeded by surface magnetic structures (Mullan \& Waldron 2006);
and models requiring magnetic fields and collisionless shocks
(Pollock 2007).

Resolved X-ray lines have served as an impetus to understand more
accurately the nature of the hot plasma component. And encoded
within these detailed X-ray emission line shapes is the required
information both about the formation process of the line (i.e., the
density and temperature which determines the emissivity) and the
vector velocity field. Although these various approaches have
certainly had successes in trying to decode these line profiles,
there remain open questions about understanding the temperature and
emission measure distributions, and the radial location of hot
plasma formation and maintenance.  In particular, previous
considerations of line profiles from clumpy/porous winds (e.g.,
Owocki \& Cohen 2006; Oskinova \etal\ 2007) have focused on issues
of clump geometry (pertaining to photon escape) and clump distributions
(pertaining volume filling factors), but these have not self-consistently
included temperature distributions implied by the structures
themselves, as for example in planar shocks.  Even smooth wind
considerations have been geometrical in nature (e.g., Owocki \&
Cohen 2001; Ignace \& Gayley 2002).  The models presented here have
the benefit of self-consistently including the detailed temperature
distribution of the shocked structures, within the context of the
assumed model.

The underlying model for clump bow shock structure was presented
in Cassinelli \etal\ (2008; hereafter Paper~I), who considered the
shape, temperature, and density of bow shocks that form around wind
clumps.  In this second paper of the series, we are explicitly
interested in the line profiles that form from these bow shock
structures and how features of the clump bow shock paradigm may
contribute to understanding the observed shapes of massive star
X-ray line profiles.  In section~\ref{sec:model} the results of
Paper~I are briefly reviewed.  In section~\ref{sec:aclump}, line
profiles are calculated and discussed for individual clump bow
shocks, emphasizing the diversity of line shapes that can result.
Section~\ref{sec:ensemble} describes line profiles that from an
ensemble of clumps, including the limitng case of many clumps and
the case of a discrete ensemble
of randomly placed clumps.  Section~\ref{sec:conc}
presents concluding remarks about these results and needed future
areas of stdy.  The Appendix details considerations of the temperature
distribution in the wind for our model prescription.

\section{Model Description}   \label{sec:model}

Our model calculations of X-ray emission line profiles produced by
a wind distribution of clumps and their associated bow shock structures
are based on simulations discussed in Paper~I.  Our
results apply to the hypersonic limit, namely that the Mach number
is high (~$\gtrsim~10$), which is an excellent description of the
situation in a massive star wind where the terminal
speed $\vinf \sim 1000$~km~s$^{-1}$
and the gas thermal speed is $\approx 100$ times smaller.  This means
that the bow shock structure -- its shape, and its relative density
and temperature distributions -- are largely independent of the actual
Mach number (e.g., Hayes \& Probstein 2004).

However, since line profile calculations require detailed information
on the actual velocity field of a large number of wind
distributed clumps, we need to establish the distribution of the
line-of-sight (LOS) velocities as seen by an observer. The simulations
of Paper~I were carried out in the rest frame of a rigid and
spherically symmetric clump where a fast moving wind with plane
parallel symmetry and constant density sweeps across the face of a
clump.  The plane-parallel approximation applies when the the radius
of the clump, $R_{\rm cl}$, is small compared to the wind clump
radial location, $r$.  To compute synthetic line profile shapes the
vector velocity field for the bow shock found in Paper~I (see their
Fig.~4), which is accomplished through {\it relative} velocity
vector defined by

\begin{equation}
{\bf \Delta V (r)} = \left[V_W (r) - V_{\rm cl}(r)\right] \,{\bf \hat{r}}
\label{eq:DV}
\end{equation}

\noindent where $V_W$ is now the preshock ambient (or inter-clump)
stellar wind speed at the site of the clump, $V_{\rm cl}$ is the
clump speed, and both are measured relative to the stellar rest
frame. We are assuming that $V_W$ and $V_{\rm cl}$ are purely radial
and thus functions only of $r$.  Note that the magnitude 
$\Delta V$ is the same in both the stellar and clump rest frames.

Our calculations apply to both cases of a clump moving faster or
slower than the ambient medium because the simulations are conducted
in the rest frame of the clump.  As long as the relative velocity
($\Delta V$) between the clump and the surrounding gas medium is
hypersonic, the same bow shock structure results.  The only practical
difference for line profiles is in which direction the bow shock
opens with respect to the star center.  If the clump is moving
radially outward faster than the wind, the stagnation point will
be ahead of the clump, and the bow shock opens toward the star.
Similar scenarios have been discussed by Guo (2010) and Waldron \&
Cassinelli (2009).  If the clump is slow, then the geometry flips
by $180^\circ$, like in the application to $\tau$~Sco for infalling
clumps (Howk \etal\ 2000).

An evaluation of the line profile shape requires knowledge of the
LOS Doppler shifts toward an observer in a specified direction.
Evaluating the Doppler shifts requires the introduction of several
coordinate systems.  We assume that the stellar wind is structured
but spherical in a time averaged sense.  With no special direction,
the stellar and observer coordinate systems are chosen to be
conincident.  Cartesian coordinates $(X,Y,Z)$ are introduced, with
associated standard spherical coordinates $(r,\vartheta,\varphi)$,
where the polar angle $\vartheta$ is measured from the $+Z$ axis.
The observer is located along the $+Z$ axis.

For each clump we adopt the cylindrical coordinate system $(\varpi,
\phi, z)$ used in Paper~I, where the $z$-axis in cylindrical
coordinates coincides with the symmetry axis of the bowshock, and
$\varpi$ is the cylindrical radius.  The clump center corresponds
to $\varpi=0$ and $z=0$.

In addition, since we envision these clumps as moving radially from
the star, the symmetry axis of the bow shock is also the radial line
from the star center to the clump center.  Hence, a key condition
inherent in these two coordinate system definitions is that ${\bf
\hat{z} \cdot \hat{r}} = \pm 1$ is maintained for all wind distributed
clumps, where the sign indicates whether the bow shock opens away
from the star ($+$) or toward the star ($-$).  

The specific points of our model are discussed in the following
sections.  We first start with a brief review of the bow shock
properties found in Paper~I as modified by using a relative velocity.

\subsection{Overview of the Bow Shock Properties}

\subsubsection{Geometry}

The numerical simulations of Paper~I showed that the 
shape of the bow shock can be well-described by the form

\begin{equation}
\frac{z - z_0}{R_{\rm cl}} = a\left(\frac{\varpi}{R_{\rm cl}}\right)^{m},
\label{eq:zGynumbs}
\end{equation}

\noindent with $a=0.35$, $m=2.34$, and $z_0 = -1.19 R_{\rm cl}$,
hence a shape not far from a parabola.  The bow shock apex forms at
a distance of $0.19 R_{\rm cl}$ above the clump surface.

In addition to the bow shock shape, it was demonstrated that the
derivative of the bow shock shape (i.e., the position-dependent tangent)
is the key parameter in determining the velocity, temperature $T$, 
and emission measure $EM$ distributions along the bow shock surface.
In Paper~I we defined this derivative as 

\begin{equation}
\frac{dz}{d\varpi} \equiv g(\varpi) = a\, m\, (\varpi/R_{\rm cl})^{m-1},
\label{eq:g}
\end{equation}

\noindent It is convenient here to introduce an angle $\alpha$
that is related to the curvature of the bow shock, with

\begin{equation}
\tan \alpha = g(\varpi).
	\label{eq:alpha}
\end{equation}

\noindent Note that in Paper~I, we had defined this angle as $A_1$
but here prefer to use $\alpha$.  

\subsubsection{The Velocity Field}

One of the major findings from Paper~I was that the $EM$ was dominated
by the immediate post-shock gas.  So an ``On The Shock'' (OTSh)
approximation was introduced, whereby the density $N$, temperature,
and velocity relevant to X-ray observables are described by conditions
along the bow shock surface, and thus by the geometry described in
the preceding section. Hence, the velocity field in the rest frame
of the clump that is needed to synthesize line profiles is known
analytically at every point along the bow shock for the known surface
geometry.

The simulation was based on the assumption that the radius of the
clump $R_{\rm cl} \ll r$, so that the incident interclump wind flow
was essentially plane parallel.  Thus ${\bf \Delta V} = \Delta
V\,\hat{z}$ in the rest frame of a clump itself (c.f., Fig.~4
of Paper~I).  We introduce the unit vectors $\hat{n}$ as the
outward normal to the shock and $\hat{l}$ as a unit vector
parallel to the shock in the direction away from the apex.  The
jump conditions for a strong oblique shock were applied to derive
the velocity components perpendicular and parallel to the shock
front.  Using primes to denote velocities in the clump rest frame,
the {\em post-shock} velocity components are given by

\begin{eqnarray}
{\bf V'}_{P,\perp} &=& -\frac{1}{4} \, |\Delta V|\,\cos\alpha \, \hat{n} \\ \nonumber
	& = & -\frac{1}{4}\,\frac{1}{\sqrt{1+g^2}}\,|\Delta V|\,\hat{n},\label{eq:vpperp}\\
{\bf V'}_{P,\parallel} &=& |\Delta V|\,\sin\alpha \,\hat{l} \\ \nonumber
	& = & \frac{g}
	{\sqrt{1+g^2}}\,|\Delta V|\,\hat{l} \label{eq:vppara}
\end{eqnarray}

\noindent where $\Delta V$ is from equation~(\ref{eq:DV}).
Note that $\Delta V$ can be positive or negative depending on whether
the clump is traveling faster or slower than the interclump wind,
but the perpendicular velocity component is always away from the
bow shock symmetry axis, and the parallel component is always
downwind of the apex position.

The postshock velocity field is a function of clump's radial distance
from the star via $\Delta V$ and also location along the bow shock
implicitly through the curvature factor $g(\varpi)$.
The total postshock flow speed{\footnote{We point out that
eq.~\ref{eq:VPsq} corrects a typo appearing in eq.~(10) of Paper~I
}} anywhere along the bow shock is

\begin{equation}
V'_P = \frac{1}{4}\, \sqrt{\frac{1+16g^2}{1+g^2}}\,\Delta V.
\label{eq:VPsq}
\end{equation}

\noindent As expected, the postshock velocity takes on a value of
$V'_P = \Delta V/4$ for a head-on collision at the bow shock apex
where $g=0$.  Far downstream of the apex, the speed approaches $V'_P
= \Delta V$ for $g \gg 1$ as the shock becomes extremely oblique.

In order to determine Doppler shifts of the X-ray emitting material,
the postshock velocity field must be evaluated in the observer
frame.  To accomplish this, it is convenient to express the vector
flow in the clump system as components of $\hat{z}$ and $\hat{\varpi}$
which can readily be transformed to the star/observer system.  This
requires a standard rotation of coordinate systems from $\hat{n}$
and $\hat{l}$ to $\hat{\varpi}$ and $\hat{z}$, which is given by

\begin{eqnarray}
\hat{n} & = & \hat{\varpi}\sin\alpha\mp \hat{z}\cos\alpha \nonumber \\
\hat{l} & = & \hat{\varpi}\cos\alpha\pm \hat{z}\sin\alpha \nonumber 
\end{eqnarray}

\noindent The distinction in sign for the $z$-component is related
to whether the clump moves faster or slower than the interclump
wind.  If slower, then the bow shock opens away from the star and the
upper sign is used; if faster, then the bow shock opens toward the
star, and the lower sign is used.

With these preceding expressions, the postshock velocity in the
rest frame of the star (unprimed system) becomes

\begin{eqnarray}
{\bf V}_P & = & {\bf V}'_P + V_{\rm cl}\,\hat{z} \nonumber \\
   & = & |\Delta V|\, \left[ \frac{3}{8}\sin 2\alpha\,
     \hat{\varpi}\pm \frac{1}{4}(1+3\sin^2 \alpha)\,\hat{z} \right] \nonumber \\
& &  + V_{\rm cl} \,\hat{z}, \\
 & \equiv & V_\varpi\,\hat{\varpi}+\left[ V_{\rm cl} + V_{\rm z}\right]\,\hat{z},
\end{eqnarray}

\noindent where $V_\varpi \ge 0$ is always true, but $V_{\rm z} \ge
0$ for the case that the clumps are slower than the interclump wind
and $V_{\rm z} \le 0$ for clumps that are faster.

The observed velocity shift for flow at a point in the
wind is $v_Z = - \hat{Z}\cdot {\boldmath V}_P$, using lowercase
``$v$'' to signify that the velocity is for the observer.  Carrying
out the dot product yields

\begin{equation}
v_Z  =  -\left[V_{\rm cl} + V_{\rm z}\right]\, \cos\vartheta
 + V_\varpi\,\cos\varphi\,\sin\vartheta .
\end{equation}

\noindent Owing to axial symmetry of the bow shock, $V_\varpi$ and
$V_{\rm z}$ are functions only of distance from the bow shock apex
along its surface.  The velocity field reduces to $V_P \approx
V_\varpi$ near the bow shock apex and $V_P \approx V_{\rm z}$ in
the far outer wings of the bow shock.

\subsubsection{The Temperature and Emission Measure Distributions}

With the OTSh approximation, the runs of $T$ and $EM$ along the
shock are monotonic functions of path length from the shock stagnation
point.  The peak temperature at the bow shock apex of an
individual clump is $T_A$ with a
value given by

\begin{eqnarray}
T_A  & =& \frac{3}{16}~ \frac{\mu m_H}{k}~\Delta V^2 \\
       & =& 14~\mbox{MK} \, \left(\frac{\mu}{0.62}\right)\,
     \left(\frac{\Delta V}{1000~\kms}\right)^2, 
\label{eq:TA}
\end{eqnarray}

\noindent where in the latter expression we have evaluated the
constants assuming a fully ionized gas with solar abundances.

The path length downwind of the apex can be expressed as a function
of impact parameter $\varpi$ for known bow shock geometry $z(\varpi)$.
The postshock temperature, $T$, along the shock is found to be

\begin{equation}
T(\varpi) = \left(\frac{1}{1+g^2}\right)~T_A .
     \label{eq:TP}
\end{equation}

\noindent The power-law form of equation~(\ref {eq:g}) provides
$g(\varpi)$, and the temperature distribution reduces to

\begin{equation}
\frac{T}{T_A} = (1~+~g^2)^{-1} = \left[~1~ + ~0.67\left(
	\frac{\varpi}{R_{\rm cl}} \right)^{2.68} \right]^{-1}
\label{eq:temp_otsh}
\end{equation}

The line profile calculation also requires an emission measure
distribution.  A result of Paper~I was that the differential emission
measure ($DEM = dEM/dT$) is a monotonic power-law function of
temperature and thus location along the bow shock with distance
from the apex.  One can conveniently parameterize the distribution
with

\begin{equation} 
\frac {dEM} {dT} = \frac{EM_\circ (r)} {T_A (r)} \,\left(\frac{T}{T_A}
	\right)^{-7/3}.
\label{eq:demdtfin}
\end{equation}

\noindent where the mapping between $\varpi$ and $T$ is made through
the factor $g$ using equation~(\ref{eq:temp_otsh}).  The emission
measure scaling parameter $EM_\circ$ is given by {\footnote{Eq.~(22)
in Paper~I is missing a factor of $(m-1)^{-1}$ which leads to a
slightly smaller scale factor for $EM_{\circ}$ as compared to the
Paper~I result.}}

\begin{equation} 
EM_{\circ} = 5.1 \times 10^{51} ~ {\rm cm}^{-3}\, \left( \frac {R_{\rm cl}} {10^{10}} 
     \right)^3 \left(\frac{N_W} {10^{10}}\right)^2 \left( \frac {\DL} {R_{\rm cl}}\right),
     \label{eq:EM0}
\end{equation}

\noindent where we have assumed a strong shock such that the postshock
number density is $N_P = 4 N_W$, where $N_W$ is the preshock
interclump wind number density, and $R_{\rm cl}$ and $N_W$ have
both been scaled to the values used in the simulation.  From Paper~I,
we found that $\DL = 0.1 R_{\rm cl}$ well approximates the simulation
results.  Our model does not predict the evolution of $R_{\rm cl}$
through the flow.  To compare most easily our results with those
of previous works, we adopt a scaling of $R_{\rm cl}^3 \propto r^2$.
Implicit then is that the emission measure varies as $EM_\circ
\propto (rV_W)^{-2}$ in form, like a smooth wind with a constant filling
factor of hot plasma (e.g., Ignace 2001).

\subsubsection{The Interclump and Clump Velocity Distributions}

The critical parameter of the bow shock model is the relative
velocity defined in equation~(\ref{eq:DV}) since it determines the
magnitude of both the postshock X-ray temperature structure and the
velocity field. Our approach is to assume a two-component
wind model where the ambient wind and clump entities follow different
velocity laws.  We adopt the commonly used standard $\beta$ velocity
law prescription defined as

\begin{equation}
V(r) = V_\infty \, \left( 1-\frac{b}{r} \right)^\beta,
\end{equation}

\noindent with $V_\infty$ the terminal speed and $b < R_\ast$ so
that the radial wind speed is non-zero at the wind base taken to
be the stellar radius $R_\ast$.  The value of $b$ sets the initial
flow speed $V_0$, with

\begin{equation}
V_0 = V_\infty\,\left(1-\frac{b}{R_\ast}\right)^\beta.
\end{equation}

\noindent In all cases considered, the $b$ parameter will be fixed
at the same value in both the interclump and clumped wind velocity
laws.  In order to limit the number of free parameters for our
exploratory investigation, we also assume that both components
achieve the same terminal speed.

The smooth interclump wind component will be described by the parameter
$\beta_W$.  Throughout this paper, we adopt $\beta_W=1$ as typical of
OB~star wind solutions.  Then $\beta_{\rm cl}$ represents possible clump
velocity distributions.  Note that $\beta_{\rm cl} > \beta_W$ implies
slow moving clumps relative to the interclump flow; $\beta_{\rm cl} <
\beta_W$ corresponds to fast moving clumps.

\subsection{Scalings from the $\beta$-Law Prescription}

The distributions of $\TA$ and the $DEM$ can be derived from the
two $\beta$ approach.  We introduce the convenient velocity
normalization $w = V_W/V_\infty$.  With the terminal speeds for the
clump and interclump flows the same, the relation between the two
components' velocity laws are $w = w_{\rm cl}^{\beta_{\rm cl}}$,
where $w_{\rm cl} = V_{\rm cl}/V_\infty$.  Thus the velocity jump
of equation~(\ref{eq:DV}) becomes

\begin{equation}
\Delta V = V_\infty\,w\,\left[1-w^{(\beta_{\rm cl}-1)}\right].
\end{equation}

\noindent This relation can be used to find $T_A(r)$, which
proceeds as follows.

Equation~(\ref{eq:TA}) along with the preceding expression gives

\begin{eqnarray}
T_A  & = & \frac{3}{16}~ \frac{\mu m_H}{k}~\Delta V^2, \\
     & = & \frac{3}{16}~ \frac{\mu
m_H}{k}~V_\infty^2\,\left\{w\,\left[1-w^{(\beta_{\rm cl}-1)}\right]\right\}^2, \\
     & \equiv & T_{\rm lim}\,\left\{w^2\,\left[1-w^{(\beta_{\rm cl}-1)}\right]^2\right\},
     \label{eq:TAofw}
\end{eqnarray}

\noindent where $T_{\rm lim}$ is implicitly defined as the highest
possible temperature in our model of outflow that occurs for
a velocity jump that is equal to the wind terminal speed.

Now the maximum hot plasma temperature $T_{\rm max}$ in the wind
model can be determined.  In the velocity coordinate of the interclump
flow, $T_{\rm max}$ is achieved at a critical value $w_{\rm c}$
as given by

\begin{equation}
w_{\rm c} = \left(\frac{1}{\beta_{\rm cl}}\right)^{1/(\beta_{\rm cl}-1)},
\end{equation}

\noindent which in the clump velocity becomes

\begin{equation}
w_{\rm cl,c} = \left(\frac{1}{\beta_{\rm cl}}\right)^{\beta_{\rm cl}/(\beta_{\rm cl}-1)}.
\end{equation}

\noindent The radial location of $T_{\rm max}$ is at a corresponding
critical radius value of $r_{\rm c}$, with

\begin{equation}
r_{\rm c} = \frac{b}{1-w_{\rm c}}.
	\label{eq:rc}
\end{equation}

\noindent The value of $T_{\rm max}$ is determined by just two
parameters:  the value of $\beta_{\rm cl}$ and the wind terminal
speed via $T_{\rm lim}$, as given by

\begin{equation}
T_{\rm max} = T_{\rm lin} \, \left(\beta_{\rm cl}-1\right)^2\,
	\beta_{\rm cl}^{-2\beta_{\rm cl}/(\beta_{\rm cl}-1)}.
\end{equation}

\begin{figure}
\plotone{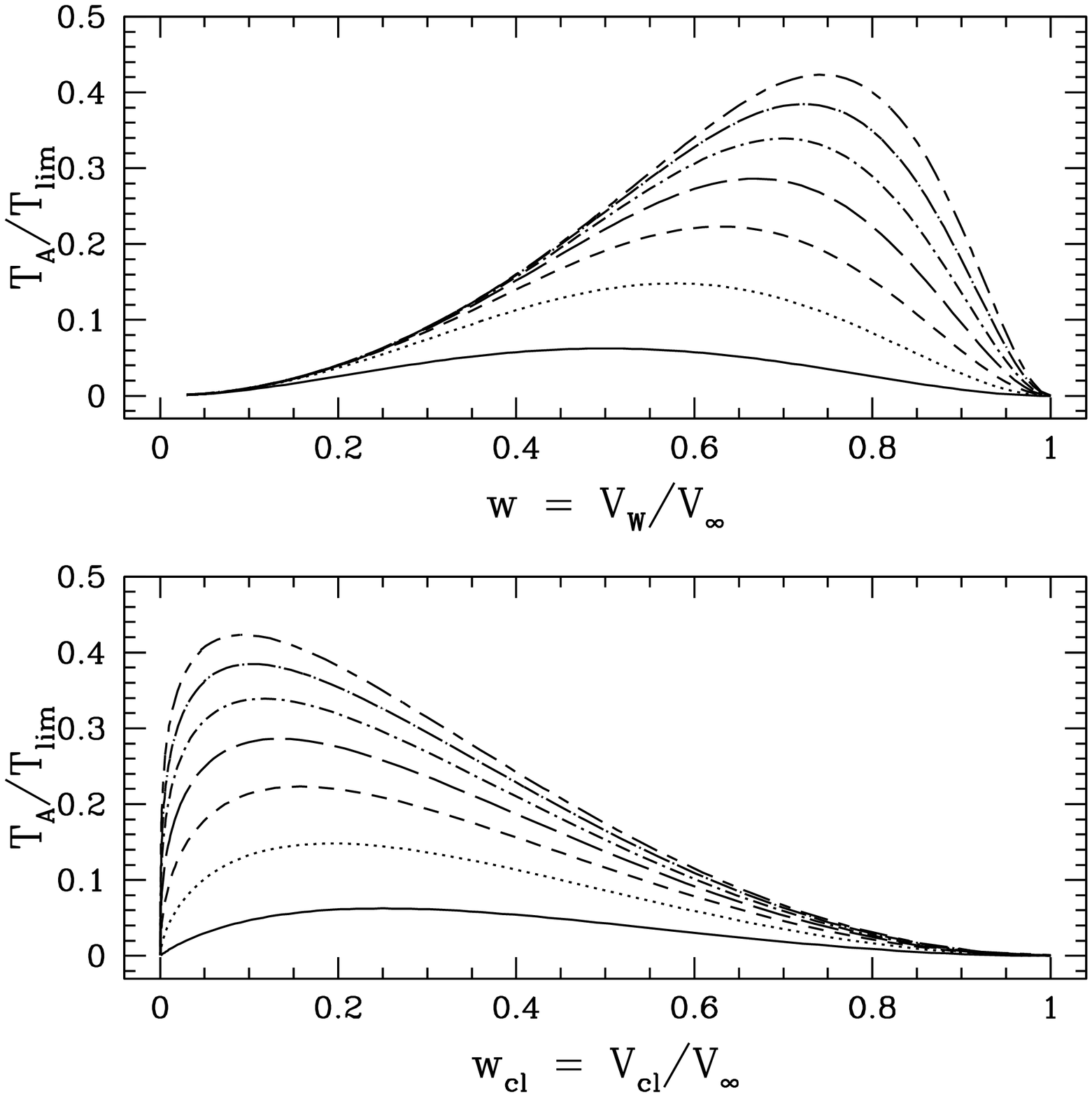}
\caption{
A plot of bow shock apex temperatures $T_A$ for clumps located at different
positions in the wind.  The temperature is normalized to $T_{\rm lim}$
(see text).  The upper panel shows location in terms of the interclump
wind velocity; lower is for the clump velocity.  Curves
are for different $\beta_{\rm cl}$ values, ranging from 2 (lowest curve) to
8 (highest curve) in integer intervals.
\label{fig1}}
\end{figure}

Figure~\ref{fig1} shows the distribution of $T_A$ in terms of the
maximum possible temperature $T_{\rm lim}$ with different curves
for different values of $\beta_{\rm cl}$.  This is
plotted against the normalized velocity of the interclump wind in
the upper panel, and against the normalized velocity of the clumps
in the lower panel.  The curves range from $\beta_{\rm cl}=2$ (lowest
curve) to $\beta_{\rm cl}=8$ (highest curve) in integer values.  As
$\beta_{\rm cl}$ increases, $T_{\rm max}$ shifts to progressively
higher velocities of the interclump wind but lower velocities for
the clump flow.  Values of $T_A$ at different velocity locations
are at the level of a tenth to a few tenths of $T_{\rm lim}$.  For
typical massive star wind speeds of 1000--3000 km s$^{-1}$, $T_{\rm
lim}$ has values of 10--100~MK.  

\section{Line Profiles for An Individual Clump}	\label{sec:aclump}

\begin{figure}
\plotone{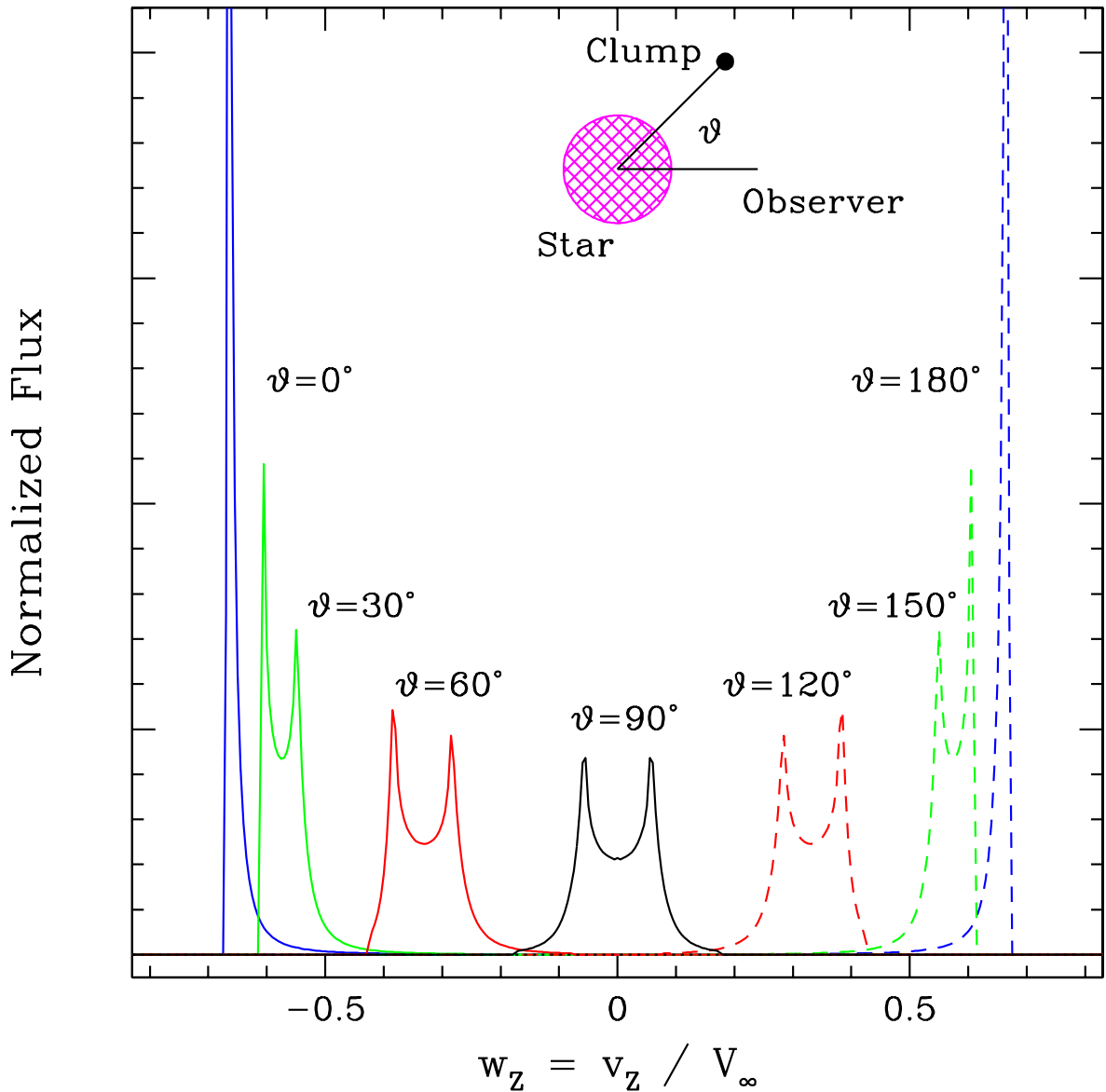}
\caption{
The inset (top center) shows the location of a clump at angle
$\vartheta$ around the star from the observer's axis.  The plot shows
example emission line profiles, all normalized to have unit area,
for individual clumps located at the indicated orientations.
In each case the clump is at the same radius, and so
all profiles have the same apex temperature $T_A$.
Solid curves are for clumps on the nearside of the star; dashed are
for ones on the far side.
\label{fig2}}
\end{figure}

Before developing emission line profiles for clumped winds, it is
instructive first to consider the emission line shape arising from
a single clump.  As an example case, we consider a clump at a
location of $2R_\ast$ that follows a $\beta_{\rm cl} =3$ velocity
law.  The velocity jump is $\Delta V \approx 0.4 V_\infty$.
Figure~\ref{fig2} demonstrates the diversity in profile shapes for
this single clump when it is located at different positions around
the star, as given by the angle $\vartheta$ illustrated by the
inset.  The abscissa is the LOS observer velocity shift $w_Z =
v_Z/V_\infty$.  Note that the profiles have been normalized to have
unit area.  Values of $\vartheta = 0^\circ, 30^\circ, 60^\circ,
90^\circ, 120^\circ, 150^\circ,$ and $180^\circ$ were considered
as labeled.  For this figure both stellar occultation and absorption
of X-rays by the clump itself are ignored, and the interclump wind
is taken to be completely optically thin to X-rays.

Except for $\vartheta = 0^\circ$ and $180^\circ$ which are clumps
that lie along the LOS to the star, the profiles tend to be
double-peaked and asymmetric.  One exception is when a clump is at
$\vartheta=90^\circ$; the profile is still double peaked but also
symmetric since it lies in the plane of the sky with the star center.
Generally, the double-horn shape is a consequence of the complex
velocity field in the bow shock.  The shapes become more nearly
single-peaked as they approach the LOS to the star center.
This is because the observer views the bow shock exactly along its
symmetry axis.  In conclusion, for a clump at $\vartheta = 0^\circ$
and $180^\circ$, only the $V_{\rm z}$ component contributes to
observed Doppler shifts, but for a clump located at $\vartheta=90^\circ$,
only the $V_\varpi$ component contributes.

For the profiles of Figure~\ref{fig2}, emission from the bow shock
contributes from the peak temperature \TA\ down to an imposed minimum
of 0.5~MK, which we use as a low temperature cut-off for hot plasma
X-ray production.  However, real lines form only over a restricted
temperature range with consequences for the line shape.  Consider
a hypothetical line that forms between 2 and 3 MK.  For a bow shock
with $\TA = 10$~MK, this line would arise spatially from an annular
band centered on the symmetry axis of the bow shock and offset from
its apex.  Consequently, realistic lines that form over different
temperature ranges will tend to have different shapes, because they
sample different portions of the postshock velocity field.

\begin{figure}
\plotone{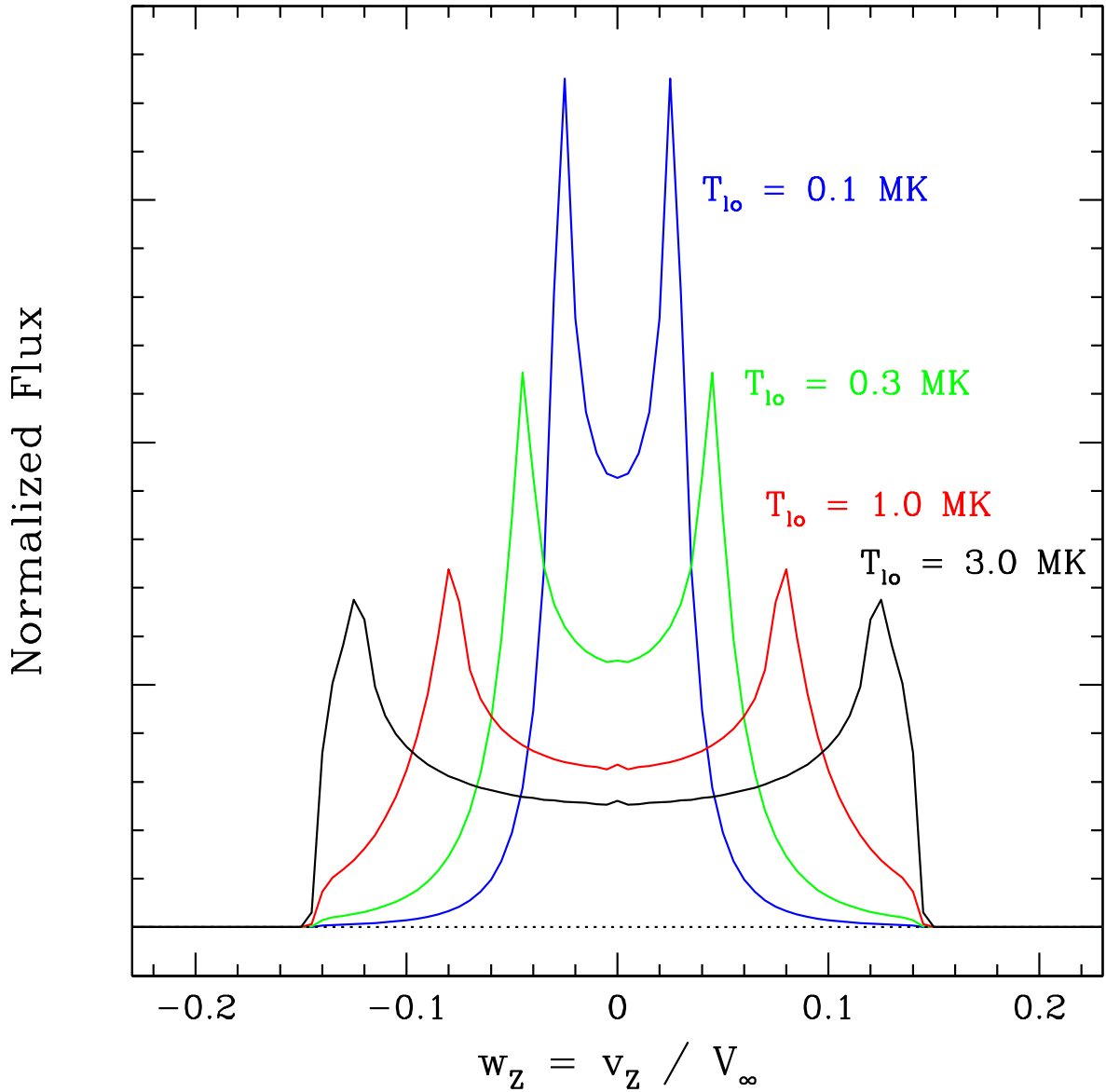}
\caption{
Similar to Fig.~\ref{fig2} but now profiles are for clumps only
at $\vartheta=90^\circ$ and with different temperature intervals.
The emissivity is taken to be constant within the temperature
range of $T_{\rm lo}$ up to $T_A$, with $T_{\rm lo}= 0.1, 0.3, 1.0,$
and 3.0~MK from the most narrow line (blue) to the broadest one (black),
respectively.
\label{fig3}}
\end{figure}

Figure~\ref{fig3} illustrates this effect through the use of simple
temperature cut-offs.  The different curves are for line emission
with different {\em low} temperature thresholds $T_{\rm lo}$.  Below
$T_{\rm lo}$ the emissivity is zero; above it the emissivity is
independent of $T$.  In this example clumps are placed at
$\vartheta=90^\circ$.  The profiles becomes progressively broader
as the lower temperature cut-off increases, with values of $T_{\rm
lo} = 0.1$~MK (blue), 0.3~MK (green), 1.0~MK (red), and 3.0~MK
(black).

To understand the growing line width with increasing $T_{\rm lo}$,
recall that the $EM$ of a clump is dominated by the low temperature
gas.  For a clump at $\vartheta=90^\circ$, the bow shock is viewed
perpendicular to its symmetry axis.  Only $V_\varpi$ components of
the postshock velocity field contribute to observed Doppler shifts.
With the lowest temperature gas found furthest downwind of the bow
apex, where the velocity vector is more nearly tangent to our LOS,
$V_\varpi$ tends to be relatively small.  Lower speed flow in the
bow shock is to be found closer to the apex; however, this flow has
a relatively larger component in the $\hat{\varpi}$ direction because
of the greater curvature, with higher LOS Doppler shifts resulting
at the bowhead.  But the bowhead is exactly where the hottest plasma
is to be found.  Thus raising $T_{\rm lo}$ means that the bowhead
region increasingly dominates the line formation, typically leading
to a broader line for the given geometry.

Two final comments.  First since increasing $T_{\rm lo}$ restricts
the contributing volume, higher values of $T_{\rm lo}$ also lead
to weaker lines for a given clump.  This is not apparent from
Figure~\ref{fig3} because each profile is normalized to unit area.
Second, the stagnation point at the bowhead is the hottest gas and
has intrinsically very low speed flow in the clump rest frame.  If
$T_{\rm lo}$ were to approach the value of \TA\, the profiles would
actually narrow, a limit not reached in the exampes of Figure~\ref{fig3}.

\section{Line Shapes from an Ensemble of Clumps}  
\label{sec:ensemble}

Although it is important to understand the emission profile from
an individual clump bow shock, stellar winds are understood to be
highly structured from many lines of evidence (e.g., Lupie \&
Nordsieck 1987; Hillier 1991; Mofatt \& Robert 1994; Lepine \&
Moffat 1999, 2008; Oskinova, Feldmeier, \& Hamann 2004; Owocki \& Cohen
2006; Prinja \& Massa 2010; Muijres \etal\ 2011).  Within our
framework, this means there is more than one clump.  Foremost is
the basic observation that X-ray emissions from single massive stars
are not highly variable.  Although there is suggestive evidence of
line variability (e.g., Nichols \etal\ 2011; Hole \& Ignace 2012),
in terms of bandpass luminosities, OB~stars are typically variable
at the level of 10\% or less (Cassinelli \& Swank 1983; Berghoefer
\& Schmitt 1994; Berghoefer \etal\ 1996, 1997).

Since we know that the observed X-ray emission from these stars
arises from a wind distribution of X-ray sources, we now need to
consider an ensemble of clump bow shocks for producing synthetic
line profiles.  We recognize the intrinsic time-dependent nature
of the problem which, in principle, requires a full radiation
hydrodynamics approach (e.g., Dessart \& Owocki 2003, 2005). Since
a goal of this paper is to present an initial analysis of line
shapes arising from bow shock structures, such a detailed approach
beyond the scope of this paper.  Our basic premise is that observed
emission lines reflect a time averaged wind flow.  High energy
resolution X-ray spectroscopy of high signal-to-noise requires
exposure times ranging from 50 to 200~ks. By contrast, the
characteristic flow time in a massive star wind is $R_\ast/v_\infty
\sim 1-10$ ks. This means that a typical massive star X-ray spectrum
is formed over multiple flow times, which tends to average over
structural variations that are stochastic.

\subsection{The Limit of Many Clumps}	\label{subsec:many}

Having considered emission profiles from an individual clump in
section~\ref{sec:aclump}, here we consider the opposite extreme of
many clumps, which we refer to as the effectively ``smooth'' limit.
It is imagined that large numbers of clumps are uniformly distributed
in radius and orientation about the star to achieve strict spherical
symmetry.  Certainly, approximate spherical symmetry is consistent
with low limits on the net continuum polarizations in O stars
(McDavid 2000; Clarke \etal\ 2002).  Polarization of unresolved
sources is related to deviations of a circumstellar envelope from
spherical (e.g., Brown \& McLean 1977; Brown, Ignace, \& Cassinelli
2000).

The idealized smooth limit has value in establishing a reference
baseline of models that can be used for interpreting line profiles
from winds with different degrees of structuring and time varying
effects.  (The latter will be treated in a separate paper.)
Additionaly, the smooth limit allows for a derivation of the DEM
from the wind as a whole.  We begin this process by referring to
equation~(\ref{eq:demdtfin}) that describes the DEM of a single
clump.  The parameters $EM_\circ$ and \TA\ are themselves functions
of clump location via equations~(\ref{eq:TA}) and (\ref{eq:EM0}).
Both of these are functions of radius (or equivalently velocity) 
alone.  Consequently, every clump in a shell of radius $r$
will have exactly the same DEM.  The global DEM for the wind will
consist of integrating contributions provided by every shell.

Working in the velocity coordinate of the interclump wind $V_W$,
the total wind DEM is given by

\begin{equation}
\left(\frac{dEM}{dT}\right)_{\rm tot} = \int_{V_1(T)}^{V_2(T)} \,
     \frac{dEM}{dT}\,\frac{dN}{dV_W}\,dV_W,
     \label{eq:windDEM}
\end{equation}

\noindent where $dN/dV$ represents the clump distribution in terms
of radial velocity\footnote{For example, Sundqvist, Owocki, \& Puls
(2011) show a clump filling factor as a function of radius, both
as inferred from observations and deduced from model simulations.
Their single peaked curve from $R_\ast$ to large $r$ corresponds
to a bell-shaped distribution in $dN/dV$ from $V_0$ to $V_\infty$.}.
In the limiting case of many clumps uniformly distributed in velocity,
$dN/dV$ is a constant.  To find the total DEM, the integration
proceeds only over shells where the temperature $T$ is high enough
to produce X-ray emission.

As described previously, there is a maximum temperature located in
the wind at radius $r_{\rm c}$ with corresponding normalized velocity
$w_{\rm c}$.  The description of apex temperatures is thus double valued
with radius.  Plus, any given shell will have a range of temperatures
from \TA\ down to a lower cut-off value.  Clearly, a particular
value of $T$ will only be found in a shell if $\TA(V_W) > T$.  It
is this condition that is used in equation~(\ref{eq:windDEM}) for
the limits of the integrand.

Using results from the preceding section and equations~(\ref{eq:TA}) and
(\ref{eq:EM0}), equation~(\ref{eq:windDEM}) becomes

\begin{eqnarray}
\left(\frac{dEM}{dT}\right)_{\rm tot} & \propto & 
     \left(\frac{T}{T_{\rm lim}}\right)^{-7/3} \, \int_{w_1}^{w_1} \,
     \left(\frac{1-w}{w}\right)^2 \\ \nonumber
 & & \times \left\{w^2\,
     \left[1-w^{(\beta_{\rm cl}-1)})\right]\right\}\,\Gamma(w)\,dw,
     \label{eq:intDEM}
\end{eqnarray}

\noindent where $w_1$ and $w_2$ represent the velocity interval
between which $T$ is achieved, and $\Gamma$ is a correction factor
for stellar occultation.  The latter is given by

\begin{equation}
\Gamma = \frac{1}{2}\,\left[ 1+\sqrt{1-\left(\frac{1-w}{b}\right)^2} \right].
\end{equation}

\noindent Note that the integral is over a fixed $T$, hence the
temperature dependence $T^{-7/3}$ can be factored out of the integral.
Thus the integral that remains represents a temperature-dependent
modification to the power law for a single clump.  The integrand
is not overly complex, but the integration limits
$w_1(T)$ and $w_2(T)$ tend not to be analytic.  (See App.~B for
solutions of $w_1$ and $w_2$ in the special cases of $\beta_{\rm
cl}=1/2, 2$, and 3.)

\begin{figure}
\plotone{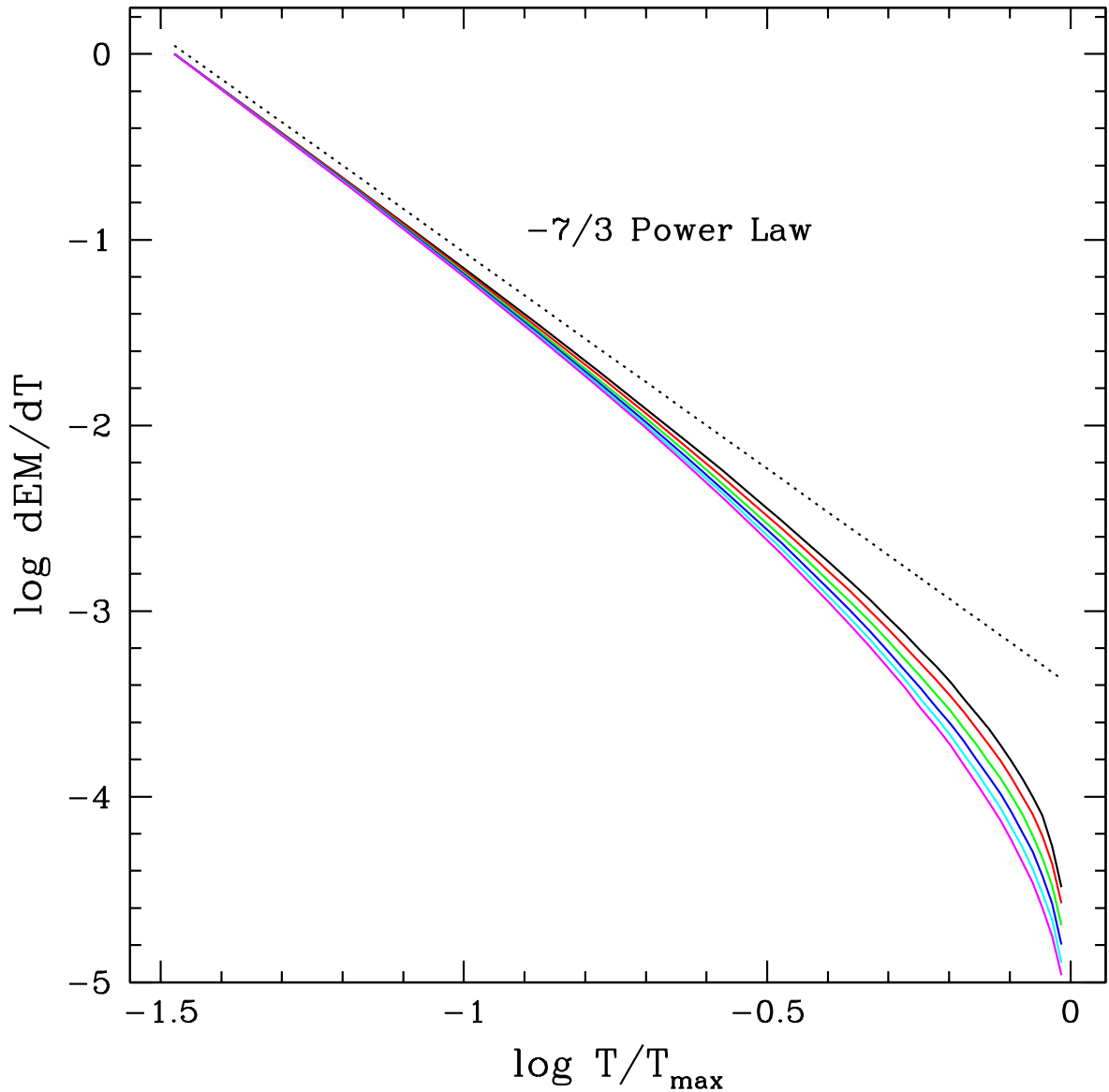}
\caption{A logarithmic plot of the intrinsic total DEM against
temperature in the smooth wind limit.  Temperature is normalized
to $T_{\rm max}$.  Curves are for $\beta_{\rm cl}$ values of
even integers between
2 to 12, inclusive.  The dotted line is for a $-7/3$ power law as
would apply to a single clump.  The curves have been shifted to
have the same value at the lowest temperature for ease of comparison.
Despite the wide range of $\beta_{\rm cl}$ values, similar
overall DEM distributions result.
\label{fig4}}
\end{figure}

Figure~\ref{fig4} displays the results of calculations for the total
DEM of the wind at even values of $\beta_{\rm cl}$ from 2 to 12.
The DEM is plotted logarithmically against $T/T_{\rm max}$.  The
curves have been shifted to a zero value at the lowest temperature
used.  The results all lie very close to each other with departures
from a $-7/3$ slope occuring only at higher values of $T$, as
emphasized by the dotted line for a power-law of $-7/3$ slope.
Different $\beta_{\rm cl}$ values yield the $-7/3$ slope at low $T$
because in our model the cooler X-ray emitting plasma is to be found
essentially throughout the wind.  The slight steepening toward
larger $T$ becomes a downturn as $T_{\rm max}$ is approached, because
only the hottest components are severely restricted in radial locale.

\begin{figure}
\plotone{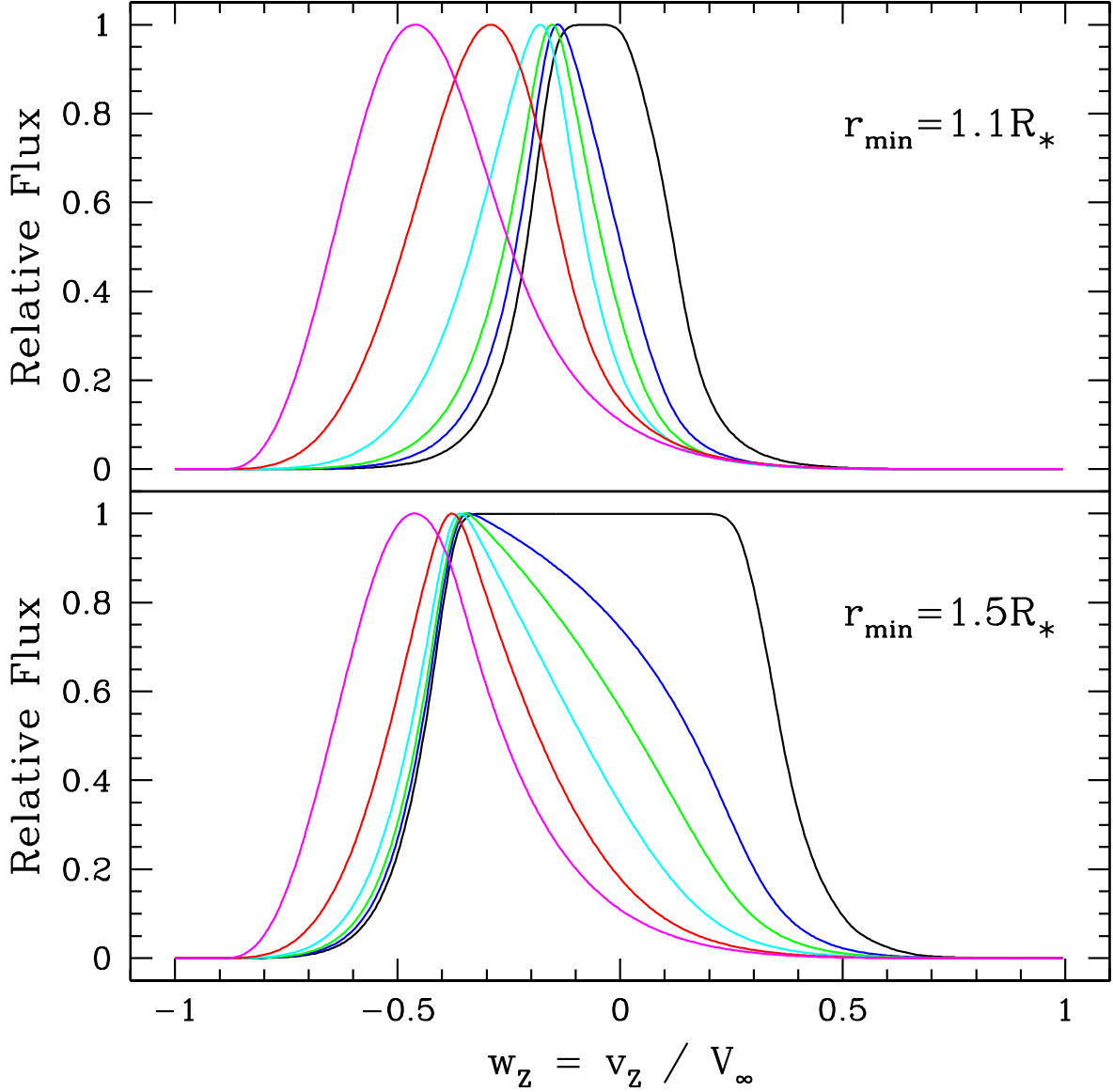}
\caption{
Illustrative emission line profiles for the smooth limiting case.
Hot plasma is assumed not to exist interior to $r_{\rm min}$, with
a value of $1.1R_\ast$ for the upper panel and $1.5R_\ast$ for the
lower one.  These models are for fast clumps with $\beta_{\rm
cl}=0.5$.  Different colored curves are for different levels of
interclump wind photoabsorption, with values of $\tau_\ast = 0,
0.5, 1, 2, 4,$ and 8 that lead to increasingly blueshifted lines.
\label{fig5}}
\end{figure}

\begin{figure}
\plotone{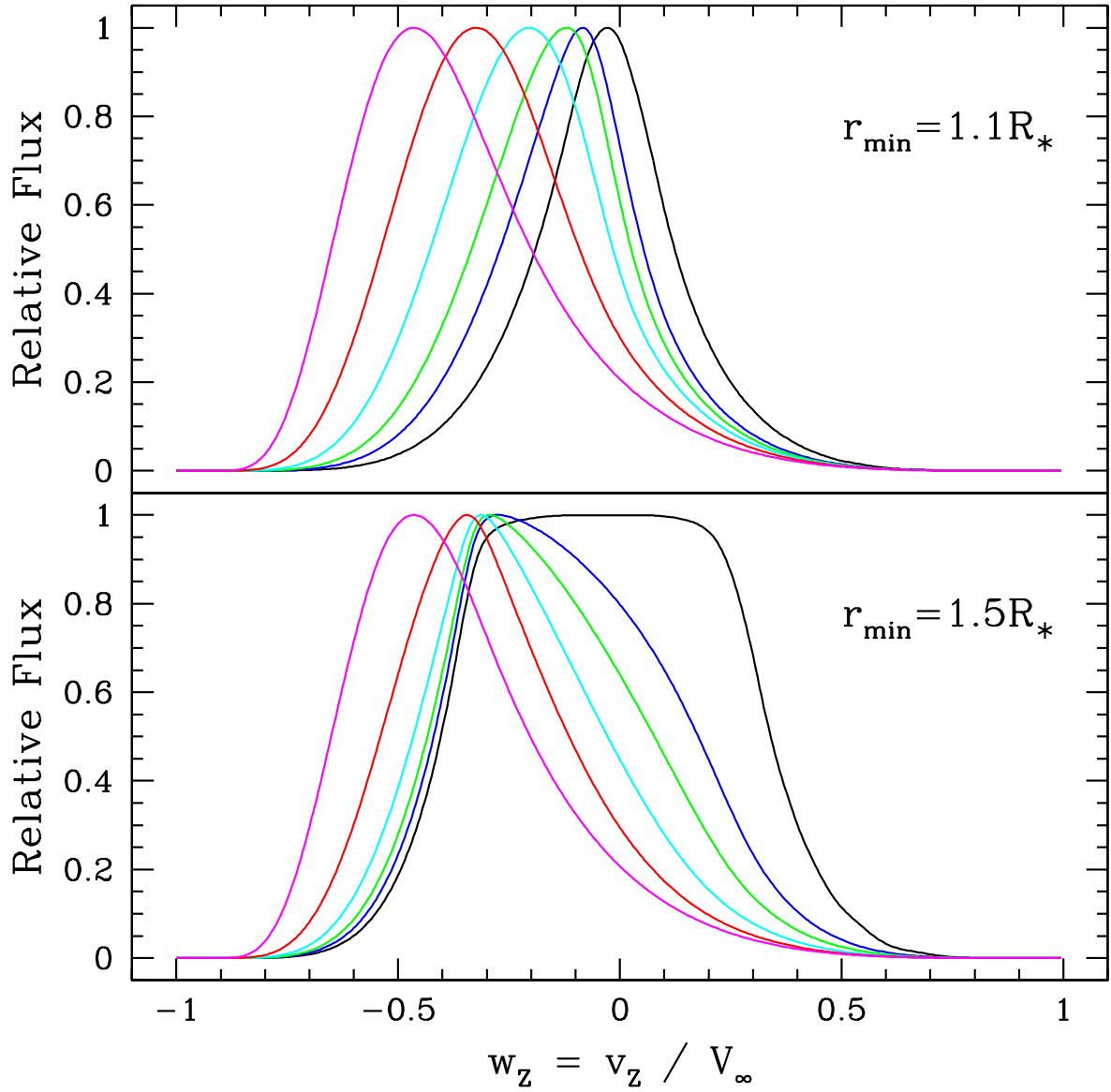}
\caption{
As in Fig.~\ref{fig5}, but now for slower moving clumps with
$\beta_{\rm cl} = 3$.
\label{fig6}}
\end{figure}

The range of line profiles that result in the smooth limiting case
are displayed in Figures~\ref{fig5} and \ref{fig6}.  A stellar wind
terminal speed of $V_\infty = 2500$~km~s$^{-1}$
is adopted, as before.  For all cases there is a minimum radius (or
velocity) in which hot X-ray emitting plasma is to be found.  For
the upper panels, the radius is $r_{\rm min} = 1.1 R_\ast$, and for
the lower panels, it is $r_{\rm min} = 1.5 R_\ast$.  The two left
panels are profiles that result for fast clumps with $\beta_{\rm
cl}=0.5$; the two right panels are for slow clumps with
$\beta_{\rm cl}=3$.

The line emission is assumed to be optically thin (hence no resonance
scattering effects).  The model line luminosity as a function
of relative velocity shift $w_Z = v_Z / V_\infty$ is calculated by

\begin{equation}
L_{\rm line} (w_Z) = \int_{w_Z}\,\Lambda(T)\,\frac{dEM}{dT}\,e^{-\tau}\,dT
\end{equation}

\noindent where $\Lambda(T)$ is the temperature dependent line
cooling function, $\tau$ is the photoabsorption optical (see below),
and the integral is carried out over the unocculted volume.  For
initial calculations we assume simply that $\Lambda(T)$ is a constant.
We also ignore the variation of ion fraction with $T$, implicit in
the DEM factor.  In effect, these illustrative model line profiles
are meant to sample the full DEM distribution.  The inclusion of
$T$-dependence for line cooling function and the effects of ionization
balance should result in a more diverse set of line profiles.

The different colored curves in Figures~\ref{fig5} and \ref{fig6}
are for different levels of wind photoabsorption.  The assumption
is that the clumps are small compared to other scales in the problem
so that the photoabsorption optical depth $\tau$ is approximated from a
LOS integration through a smooth interclump wind.  As such, the
photoabsorption optical depth to the clump position is the same for
all points along the bow shock.  

The optical depth is calculated
following Ignace (2001), with

\begin{equation}
\tau(r,\vartheta) = \frac{\tau_\ast}{r\,\sin\vartheta}\,
     \int_0^{\vartheta} \, \frac{d\vartheta'}{1-[(b\sin\vartheta')
     /(r\sin \vartheta)]},
\end{equation}

\noindent where $\tau_{\ast}$ is the optical depth scale to the
base of the wind at $R_\ast$.  Generally, this scale is related to
the wind mass-loss rate, abundances, and energy of the particular
line transition in question.

In Figures~\ref{fig5} and \ref{fig6} profiles for values of $\tau_\ast
= 0, 0.5, 1, 2, 4,$ and 8 are calculated.  The effect of increasing
$\tau_\ast$ is to make the profiles increasingly asymmetric with
emission peaks of progressively higher blueshifts.  The black profile
is for a line with $\tau_\ast =0$; magneta corresponds to the case
of $\tau_\ast=8$.  Note that with no photoabsorption, the black
curve displays a central ``flat-top'' indicative of $V_{\rm min} =
V(r_{\rm min})$, modulo the effect of stellar occultation.

For the range of photoabsorptive optical depths used, the blueshifted
peaks all lie below about half of terminal speed.  The line widths
actually decrease slightly for low $\tau_{\ast}$, but then increase
with larger values of $\tau_{\ast}$.  Ultimately, at large optical
depths, the blueshift of the peak emission and the line width are not
sensitive to the value of $r_{\rm min}$.

\subsection{The Case of Discrete Clumps}  \label{sec:discrete}

Relaxing the assumption of a smooth distribution of clumps, a
discretely structured flow is now considered.  Models are based on
a random number generator to place clumps throughout the wind from
which the X-ray emission line profiles are computed.  Let $s$ be a
random number in the range of 0 to 1.  Individual clumps are sprinkled
in a uniformly random way about the star.  The $i$th clump will
have angular coordinates given by $\varphi_{\rm i}=2\pi s$ and
$\mu_{\rm i} = -1+2s$, where $\mu_{\rm i} = \cos \vartheta_{\rm
i}$, and of course two distinct random numbers $s$ are used to set
the two seperate coordinates for a given clump.

Radial placement requires a different approach.  In our model clumps
exist exterior to the photospheric level, $R_\ast$, out to infinite
distance, in principle.  The 1D radiative hydrodynamic simulations indicate that
the formation of strong shocks occurs primarily at low and intermediate
radii in the flow (e.g., Feldmeier, Puls, \& Pauldrach 1997),
basically where the velocity gradient is reasonably strong. Structure
can persist and evolve out to fairly large radius.  As a way of
capturing the flavor of this scenario, we choose to space clumps
such that they are statistically uniformly distributed in radial
velocity $V_W$.  In relation to the preceding section, this means
that $dN/dV$ now becomes a uniform probability distribution to be
sampled in the range of $V_0$ to $V_\infty$.

The relationship between a random number $s$ and the corresponding
velocity for that value is given by:

\begin{equation}
s = \frac{V_W-V_0}{V_\infty - V_0}.
\end{equation}

\noindent Naturally, this distribution is highly non-uniform in
radius.  On average half of the clumps lie at $V_W>0.5V_\infty$,
and the other half lie below that speed.  For $\beta_W=1$, this
means that half the clumps lie beyond $r=2R_\ast$, and half lie
interior.  One can easily incorporate different
distributions $dN/dV$, either as an exploration of parameter space
or to match known clumping properties of a particular source.  The
manner in which X-ray producing structures are placed in velocity
space influences the line profile shape.  Our choice of $dN/dV$ as
uniform is merely a convenience for purposes of illustration.

\begin{figure}
\plotone{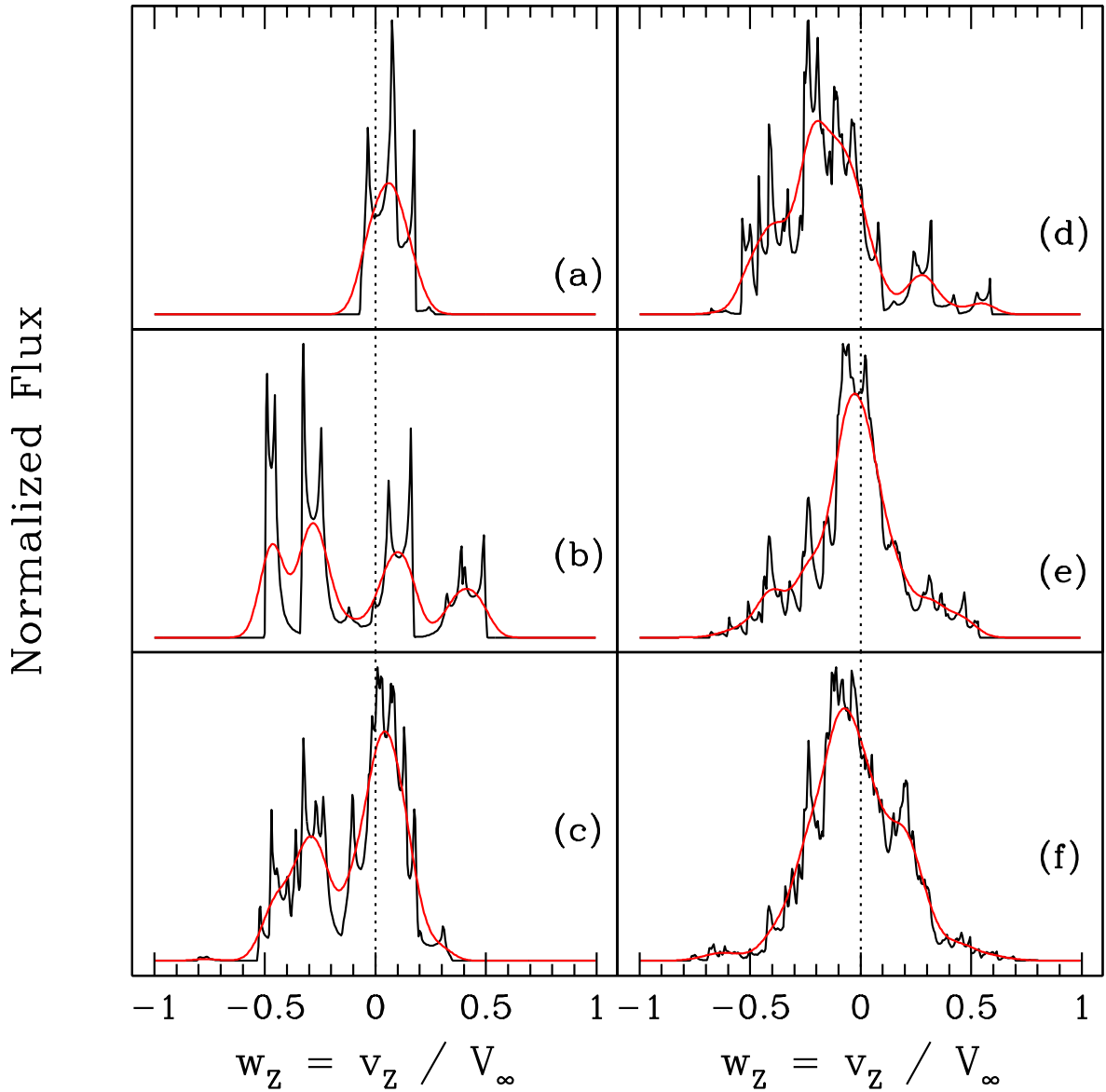}
\caption{
Line profile shapes for an ensemble of clumps with 
$\tau_\ast =0.1$.  Panels are distinguished by the number of clumps
${\cal N}_{\rm cl}$ used in the model, with (a) 4, (b) 8, (c) 16, (d) 32,
(e) 64, and (f) 128 clumps.  Model line profiles are shown in black;
overplotted are red curves that include the effects of instrumental
smearing are included.  Finite spectral resolution is approximated by
convolving model lines with a Gaussian that has $\sigma =
0.05 V_\infty$.
\label{fig7}}
\end{figure}

Figure~\ref{fig7} shows examples of emission lines for different
clump ensembles.  The wind photoabsorption optical depth is set to
a low value of $\tau_{\ast} = 0.1$.  All profiles have been normalized
to unit area and so no vertical scale of flux is provided.  The six
panels labelled (a)--(f) correspond to different numbers of clumps
${\cal N}_{\rm cl}$ with 4 in (a), 8 in (b), 16 in (c), 32 in (d), 64 in
(e), and 128 in (f).  The black curves are the intrinsic profiles
of the model calculation, whereas the red curves are convolved by
a Gaussian to simulate the effect of instrumental smearing from
finite spectral resolution.

It is important to note that the number of clumps contributing to
a given profile is generally less than the value of ${\cal N}_{\rm
cl}$.  This occurs for a couple of reasons.  First, we adopt a
threshold temperature of 0.5~MK for gas to contribute to the line.
If the apex value $T_A$ is less than the threshold, then all the
gas in the bow shock of that clump is also less than the threshold.
The threshold eliminates those clumps that are very near the
photosphere and very far away, where $\Delta V$ is too
small to generate the requisite temperatures for X-ray emission.
The second reason is that some clumps are occulted.  

With ${\cal N}_{\rm cl}$ on the order of several tens and higher, the
convolved profiles are reasonably symmetric (but not exactly so).
Of course the extent of blueshifted peak emission and line width
is a function of photoabsorption optical depth.

We have not properly dealt with the fact that there is generally a
broad range of temperatures across the bow shock.   The emission
lines of Figure~\ref{fig7} still adopt a temperature independent
line emissivity as was used for the effectively smooth wind case
of section~\ref{subsec:many}.  A temperature dependent emissivity
should be included when fitting observed line profiles for specific
sources.

\section{Summary and Conclusions}	\label{sec:conc}

Paper~I of this series presented results of a hydrodynamic simulation
for purely adiabatic cooling with a plane-parallel hypersonic flow
impinging upon a rigid spherical obstacle in the rest frame of that
obstacle.  The simulation was conducted under the assumption that
individual clump structures are much smaller than the radius at
which they are located.  In that paper the flow and temperature
structure were described, and two quite interesting simplifications
were emphasized.  First, it was found that the DEM followed a
power-law form.  Second, the emission measure was to be found
primarily in a thin ``sheath'' of postshock volume.  Thus Cassinelli
\etal\ (2008) introduced the on-the-shock approximation, or ``OTSh'',
whereby the bow shock geometry determines the $T$ and $DEM$
distributions necessary for computing observables.

In this second paper, we adopt the OTSh to model X-ray emission
lines that would arise from an individual bow shock and from an
ensemble of bow shocks.  This follows on a long string of papers
to explain the unexpected observed X-ray line profile shapes from
a number of massive stars in terms of structured flows, based on
fragments of planar shocks (Oskinova \etal\ 2004) or porosity
arguments (Owocki \& Cohen 2006).

An individual clump tends to produce an asymmetric double-horned
emission profile that is offset from line center, depending on its
radial and lateral location around the star from the observer.
Evidence indicates that massive star winds are characterized
by large numbers of clump structures.  To model the line shapes
from an ensemble of clumps, we adopted a parametric two-component
flow approach using two wind $\beta$-laws:  one for the
interclump wind flow and one for the clump flow.  The distinction
in $\beta$-laws leads to radius-dependent velocity jumps that govern
the temperature range of the bow shocks.  Of particular interest
is that this approach yields a number of semi-analytic relationships
for the $T$ and DEM distributions throughout the flow, which in
principle are properties that can be tested against observations
(e.g., Waldron \& Cassinelli 2009; Guo 2010);

Using this construction, emission line profiles were calculated in
the ``smooth'' limit of many uniformly distributed clumps and for
the case of a discretely structured flow.  As expected, peak emission
of the lines are a function of the degree of photoabsorption.  The
bow shock paradigm yields line shapes that are somewhat symmetric
at modest photoabsorption optical depths of a few, where the
influence of $r_{\rm min}$ on the line shape can no longer 
be perceived.  In contrast
to a uniform distribution of clumps, the discrete case leads to
profiles with spikey features; however, these are much too narrow
to actually resolve with current instrumentation.  Using a simple
temperature cut-off approach, we also find that profile widths can
depend on the temperature interval of line formation.  

All of these results represent a promising starting point for
tailored analyses of individual objects, for calculating spectral
energy distributions, and for investigating X-ray variability.
Previous efforts have focused primarily on geometrical considerations
for explaining X-ray line profiles shapes observed from OB stars,
in the form of discrete clumps, clump distributions, and/or filling
factor considerations.  Our results explicitly include temperature
distributions throughout the wind flow, which is a forward step
in X-ray line profile synthesis modeling.

In closing we remind the reader that our approach has relied on
simulations that adopt purely adiabatic cooling for the bow shocks.
We have begun new simulations of clump bow shocks that include
radiative cooling.  The advantage of adiabatic cooling and hypersonic
flow is that the flow geometry is independent of the Mach number.
In situations where radiative cooling is needed, the results will
depend on the density and the apex temperature achieved.
Consequently, the bow shock structure will no longer have a
``universal'' form; thus, greater complexity is the cost of greater
realism.  Preliminary results with radiative cooling suggest that
the power-law DEM in temperature derived in Paper~I persists at the
hottest temperatures, but shows a flattening toward cooler temperature
gas where radiative cooling dominates.  In the future we will
include the results of these new simulations along with realistic
temperature-dependent line emissivities to fit the line profiles
of high resolution X-ray lines from massive star winds and
to study time variable effects of X-ray emissions.

\acknowledgments The authors express appreciation for helpful comments
made by an anonymous referee.  RI, WLW, and JPC gratefully acknowledge
funding support for this work from a NASA ATFP award NNH09CF39C.
AB was funded by a partnership between the National Science Foundation
(NSF AST-0552798), Research Experiences for Undergraduates (REU), and
the Department of Defense (DoD) ASSURE (Awards to Stimulate and Support
Undergraduate Research Experiences) programs.

\appendix

\section{Appendix:  Temperature Intervals for $\beta_{\rm cl}=1/2$, 2, and 3}
	\label{sec:app}

To calculate the total DEM from a uniform distribution of many
clumps in a wind, it is necessary to find the integration limits
$w_1(T)$ and $w_2(T)$ in eq.~(\ref{eq:intDEM}).  Based on the
preceding section, this amounts to a root finding exercise involving
the following relation (see eq.~[\ref{eq:TAofw}]):

\begin{equation}
w^{\beta_{\rm cl}}-w+\sqrt{t} = 0,
\end{equation}

\noindent where $t=T/T_{\rm lim}$ and $\beta_W=1$ is assumed.  The
function is double-valued for all $\beta_{\rm cl}\ne 1$.  Note that
the clump $\beta_{\rm cl}$ can be larger or smaller than the
interclump value.  
Here solutions are given for three cases where the roots are analytic
or semi-analytic.

\subsection{Case of $\beta_{\rm cl}=2$}

The equation to be solved is

\begin{equation}
w^2-w+\sqrt{t} = 0.
\end{equation}

\noindent The roots have with values of

\begin{equation}
w_{1,2} = \frac{1}{2} \mp \frac{1}{2}\,\sqrt{1-4\sqrt{t}}.
\end{equation}

\noindent The maximum temperature occurs at $t_{\rm max}=1/16$ for which
$w_1=w_2=0.5$.  

\subsection{Case of $\beta_{\rm cl}=1/2$}

The equation to be solved is

\begin{equation}
w-\sqrt{w}+\sqrt{t} = 0.
\end{equation}

\noindent With the change of variable $x^2=w$, the condition can be
recast as

\begin{equation}
x^2-x+\sqrt{t} = 0,
\end{equation}

\noindent which is the same quadratic expression for $\beta_{\rm cl}=2$.
The roots $w_{1,2}$ for the case $\beta_{\rm cl}=1/2$ are simply the square
roots of the solutions from the $\beta_{\rm cl}=2$ case.  The maximum
temperature still occurs at $t_{\rm max}=1/16$, which in velocity is
now $w_1=w_2=0.25$.  

\subsection{Case of $\beta_{\rm cl}=3$}

The expression to be solved is

\begin{equation}
w^3-w+\sqrt{t} = 0.
\end{equation}

\noindent This cubic has three real roots; however,
one of those is negative and not physical.  There are standard forms for
the roots; here we use the trigonometric version.  An
angle $\gamma$ is introduced as defined by

\begin{equation}
\cos \gamma = - \sqrt{t/t_{\rm max}},
\end{equation}

\noindent where $t_{\rm max}=4/27$.  Then the roots become

\begin{equation}
w_2 = \frac{2}{\sqrt{3}}\,\cos(\gamma/3),
\end{equation}

\noindent and

\begin{equation}
w_1 = \sqrt{1-\frac{3}{4}\,w_2^2} - \frac{1}{2}\,w_2.
\end{equation}

\end{document}